\documentclass[floatfix,twocolumn,showpacs,prl,amsmath]{revtex4}
\usepackage{graphicx}
\usepackage{bm}
\usepackage{amssymb}
\usepackage{dcolumn}
\usepackage{subfigure}

\begin{document}
\newcommand{\vn}[1]{{\bf{#1}}}
\newcommand{\vht}[1]{{\boldsymbol{#1}}}
\newcommand{\matn}[1]{{\bf{#1}}}
\newcommand{\matnht}[1]{{\boldsymbol{#1}}}
\newcommand{\bege}{\begin{equation}}
\newcommand{\ee}{\end{equation}}
\newcommand{\bal}{\begin{aligned}}
\newcommand{\defbar}{\overline}
\newcommand{\SM}{\scriptstyle}
\newcommand{\eal}{\end{aligned}}
\newcommand{\udot}{\overset{.}{u}}
\newcommand{\exponential}[1]{{\exp(#1)}}
\newcommand{\phandot}[1]{\overset{\phantom{.}}{#1}}
\newcommand{\phandag}{\phantom{\dagger}}
\newcommand{\Trace}{\text{Tr}}
\title{Anisotropic spin Hall effect from first principles}
\author{Frank Freimuth}
\author{Stefan Bl\"ugel}
\author{Yuriy Mokrousov}
\affiliation{Institut f\"ur Festk\"orperforschung and
Institute for Advanced Simulation,
Forschungszentrum J\"ulich and JARA, 52425 J\"ulich, Germany}
\date{\today}
\begin{abstract}
We report on first principles calculations of the anisotropy of the 
intrinsic spin Hall conductivity (SHC) in nonmagnetic hcp metals 
and in antiferromagnetic Cr.
For most of the metals of this study we find large anisotropies.  
We derive the general relation between the SHC vector
and the direction of spin polarization and discuss its 
consequences
for hcp metals. Especially, it is predicted that for systems
where the SHC changes sign due to the anisotropy the
spin Hall effect may be tuned such that the
spin polarization is parallel either to the electric field or 
to the spin current.

\end{abstract}

\pacs{72.25.Ba, 72.15.Eb, 71.70.Ej}

\maketitle

Despite its relative smallness, the spin-orbit interaction (SOI)
in solids gives rise to many phenomena of technological
relevance and general scientific interest -- well-known examples
are
magnetocrystalline anisotropy and  
anisotropic magnetoresistance.
The anomalous part of the 
Hall effect~\cite{nagaosa_sinova_onoda_macdonald_ong_2010}, which is observed in
ferromagnets even in the absence of a magnetic induction field,
results from the 
spin-dependent transverse velocities which charge carriers acquire
in the presence of a longitudinal 
electric field due to SOI.
In paramagnets the spin-dependent transverse velocities of 
spin-up and spin-down electrons are exactly opposite 
generating a transverse pure spin current, which is known as
spin Hall effect 
(SHE)~\cite{dyakonov_perel_1971b,hirsch_1999,murakami_2003,sinova_2004}. 
From the theoretical point of view
SHE and anomalous Hall effect (AHE) are
thus intimately related and new insights into one of the two effects
usually improve understanding of the other. 

While the SHE had
been predicted theoretically already in 
1971~\cite{dyakonov_perel_1971b}
it was demonstrated experimentally
for the first time in 2004~\cite{kato_2004}. Since then the 
enthusiasm about the SHE
has not abated. 
It has been studied experimentally in 
semiconductors~\cite{kato_2004,sih_2005,wunderlich_2005,stern_2006}
and metals~\cite{valenzuela_tinkham_2006}.
As the SHE allows to access the 
spin degree of freedom
of the electron without making use of magnetism it is believed
to play an important role in future generations of spintronic devices. 

It is well known both from theory
and experiment (see Ref.~\cite{roman_mokrousov_souza_2009} and references therein)
that the 
anomalous Hall conductivity exhibits anisotropy, i.e., it is 
dependent on the orientation of 
magnetization $\vn{M}$.
Experimental evidence for the anisotropy of the SHE has been reported for
AlGaAs quantum wells~\cite{sih_2005}.
In contrast to the AHE the SHE is isotropic in cubic materials, i.e., in 
order to observe the anisotropy of the SHE non-cubic 
materials have to be considered~\cite{anishenoncub_chudnovsky}.
Besides bearing potential for applications the anisotropic Hall effects
are also interesting from the point of view of information encoded in them
about the Fermi surfaces and mean free paths of metals. 
Furthermore, it has been proposed~\cite{anishenoncub_chudnovsky} to 
exploit the anisotropy 
in order to distinguish experimentally between inverse spin Hall effect 
and competing effects caused by the magnetic field of the transport
current. 
So far, \textit{ab initio} calculations 
of the anisotropy of SHE have not been discussed in the literature.

In the present work we undertake a detailed study of the anisotropy of the
SHE in nonmagnetic hcp metals and in antiferromagnetic Cr. In particular, 
the general expression for the orientational
dependence of the SHC in hcp and tetragonal metals is derived and the
anisotropies are calculated from \textit{ab initio} 
within the density functional theory.
Generally, in antiferromagnets the SHE
is expected to allow the generation
of pure spin currents like in paramagnets. While the AHE has recently
been studied in complex magnetic
structures~\cite{skyrmions_AHE_DM_spiral_magnet_nagaosa_2009} 
first principles calculations of the SHE have been limited
to paramagnets so far. 
Since the magnetic structure breaks the cubic symmetry   
antiferromagnets such as Cr always 
exhibit anisotropic SHE.

The anisotropy of the AHE manifests itself in the dependency of 
the magnitude of the
conductivity vector on the magnetization direction. 
The conductivity vector $\vht{\sigma}^{\text{AHE}}(\vn{M})$ relates the 
anomalous Hall current density $\vn{J}^{\text{AHE}}$ to the electric   
field:
\bege\label{eq_define_anomalous_hall_conductivity_vector}
\vn{J}^{\text{AHE}}=\vn{E}\times\vht{\sigma}^{\text{AHE}}(\vn{M}).
\ee
While the anomalous Hall current is always perpendicular to the
electric field, it is not necessarily perpendicular to the magnetization,
since conductivity and magnetization vectors are not parallel in general.
In the case of the SHE in paramagnets there is no
magnetization vector $\vn{M}$ to control, only the direction of the
applied electric field can be varied. However, the spin polarization
of the induced spin current depends on the direction in which the
spin current is measured (see Fig.~1(a)). Hence, for a fixed electric field 
a given spin polarization $\hat{\vn{S}}$
is measured only in a certain direction. 
Thus, in analogy to
Eq.~\eqref{eq_define_anomalous_hall_conductivity_vector} we may write
\bege\label{eq_define_spin_hall_conductivity_vector}
\vn{Q}^{\hat{\vn{S}}}=\vn{E}\times\vht{\sigma}(\hat{\vn{S}}),
\ee  
where $\vn{Q}^{\hat{\vn{S}}}$ is the spin current density 
and $\vht{\sigma}(\hat{\vn{S}})$ is the SHC vector.
If the magnitude of the SHC vector depends on the
spin polarization direction $\hat{\vn{S}}$ the SHE is said to be anisotropic.

The spin current is characterized by 
velocity and spin polarization.
Hence, the spin current density $\vn{Q}$ is a tensor in 
the 9-dimensional space $\mathbb{R}^{3}\otimes\mathbb{R}^{3}$
spanned by the basis vectors $\hat{\vn{e}}_{i}\otimes\hat{\vn{s}}_{s}$. 
For clarity we use the 
symbols $\hat{\vn{s}}_{x},\hat{\vn{s}}_{y}$ and $\hat{\vn{s}}_{z}$ 
to denote the unit vectors of spin polarization while
$\hat{\vn{e}}_{x}$, $\hat{\vn{e}}_{y}$ and $\hat{\vn{e}}_{z}$ are 
the unit vectors of velocity.
In addition to the SHC 
vector we define the
tensor of SHC $\sigma_{ij}^{s}$, which has 
three indices: $i$ denotes the direction of spin current,
$j$ the direction of applied external electric field, and
$s$ the direction of spin polarization of the spin current. 
The general expression for the linear response of the 
spin current density to an applied external electric field
is given by
\bege\label{eq_spincurrent_compact}
\vn{Q}=
\sum_{ijs}\sigma_{ij}^{s}\hat{\vn{e}}_{i}\otimes\hat{\vn{s}}_{s}E_{j}.
\ee
Comparing Eq.~\eqref{eq_define_spin_hall_conductivity_vector} and 
Eq.~\eqref{eq_spincurrent_compact} we find that the
SHC vector $\vht{\sigma}(\hat{\vn{S}})$
and the SHC $\sigma_{ij}^{s}$
are related as follows:
\bege\label{eq_relation_sigma_sigmavector}
\sigma_{k}(\hat{\vn{S}})=\frac{1}{2}\sum_{ijs}\epsilon_{ijk}\sigma_{ij}^{s}S_{s},
\ee
where $\sigma_{k}$ is the $k$-th component of the conductivity vector,
$\hat{\vn{S}}=(S_{x},S_{y},S_{z})^{T}$ and $\epsilon_{ijs}$ is 
the Levy-Civita symbol. 
Eqns.~(\ref{eq_spincurrent_compact}-\ref{eq_relation_sigma_sigmavector})
prove Eq.~\eqref{eq_define_spin_hall_conductivity_vector}, which we 
conjectured above from 
analogy to Eq.~\eqref{eq_define_anomalous_hall_conductivity_vector}.

In cubic systems symmetry requires 
that $\sigma_{ij}^{s}=\sigma_{xy}^{z}\epsilon_{ijs}$. 
Thus, the SHC may be expressed in terms of
one material parameter, 
Eq.~\eqref{eq_define_spin_hall_conductivity_vector} simplifies into
$\vn{Q}^{\hat{\vn{S}}}=\sigma_{xy}^{z}\vn{E}\times\hat{\vn{S}}$, and the conductivity vector is
$\vht{\sigma}(\hat{\vn{S}})=\sigma_{xy}^{z}\hat{\vn{S}}$. Since the magnitude of the 
conductivity vector, $\sigma_{xy}^{z}$, is independent of $\hat{\vn{S}}$,
the SHE is isotropic in cubic systems.
The relationship between the direction of spin current and
the direction of spin polarization in cubic systems 
is illustrated in Fig.~1(a).
\begin{figure}
\includegraphics[width=8.5cm]{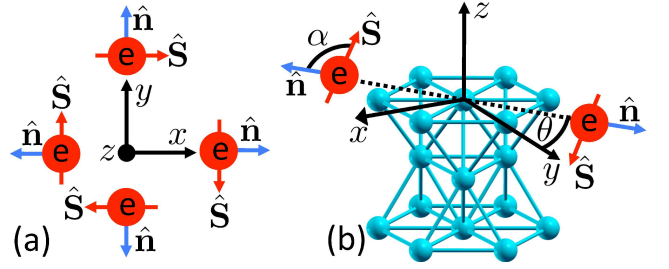}
\caption{\label{fig_she_tensor} (a) Spin currents in cubic systems induced by
an electric field in $z$-direction. For electrons \textbf{e} moving in 
$y$-direction the
spin polarization $\hat{\vn{S}}$ points in $x$-direction, while for electrons going in $x$-direction
the spin points in minus $y$-direction.
(b) Hexagonal hcp structure of the transition metal Ti. 
Due to the electric field in $x$-direction a 
spin current flows along $\hat{\vn{n}}=(0,\cos\theta,\sin\theta)^{T}$. 
In general $\hat{\vn{n}}$ and $\hat{\vn{S}}$ 
enclose an 
angle $\alpha\neq 90^{\circ}$.
}
\end{figure}

\begin{figure*}[Ht!]
\centering
\includegraphics[width=17.5cm]{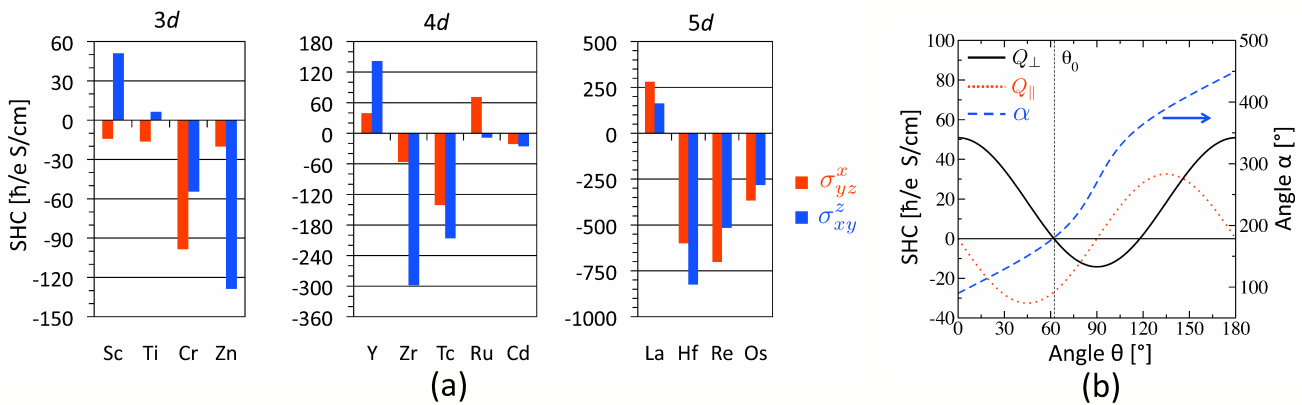}
\caption{\label{conductivies_anisotropic_she} 
(a) For the hcp metals
Sc, Ti, Zn, Y, Zr, Tc, Ru, Cd, La, Hf, Re and Os and for
antiferromagnetic
Cr the spin Hall conductivities 
$\sigma^{x}_{yz}$ and $\sigma^{z}_{xy}$ are shown as light (red) and
dark (blue) bars, respectively.
(b) Decomposition of the SHC of Sc into perpendicular and
parallel components following Eq.~\eqref{eq_q_perp}. The angle
$\alpha$ enclosed by the direction of the spin current and the 
direction of the spin polarization is also shown. At the 
angle $\theta_{0}$=62.2$^{\circ}$ the component of the spin polarization
perpendicular to the spin current vanishes and $\alpha$=180$^{\circ}$.
}
\end{figure*}

In contrast to cubic systems the SHE in hexagonal systems is anisotropic. 
Consider the structure
of the hcp transition metals,
which is illustrated in Fig.~1(b). If the electric field is applied along the
$x$-direction, the magnitude of the spin current in $y$-direction will generally differ
from the one in $z$-direction since the $x$-axis exhibits only 2-fold rotational symmetry.
The spin current density in direction $\hat{\vn{n}}=(0,\cos\theta,\sin\theta)^{T}$ is
\bege\label{eq_e_field_along_x}
\hat{\vn{n}}\cdot\vn{Q}=
-(
\sigma^{z}_{xy}\hat{\vn{s}}_{z}\cos\theta-
\sigma^{y}_{zx}\hat{\vn{s}}_{y}\sin\theta
)E_{x}.
\ee
Note that according to Eq.~\eqref{eq_spincurrent_compact} $\hat{\vn{n}}\cdot\vn{Q}$ is a 
vector pointing in the direction of spin polarization. 
We define the anisotropy of the SHE for spin polarization 
in the $yz$-plane as $\Delta_{zy}=\sigma_{xy}^{z}-\sigma_{zx}^{y}$.
For a general angle $\theta$ the components of
the spin current with spin polarization parallel 
to $\hat{\vn{n}}$ ($Q_{\Vert}$) and spin polarization perpendicular 
to $\hat{\vn{n}}$ ($Q_{\perp}$) are
given by
\bege
\begin{aligned}
\label{eq_q_perp}
Q_{\Vert}&=\hat{\vn{n}}\cdot \vn{Q}\cdot\hat{\vn{n}}=
-\frac{1}{2}\Delta_{zy}\sin(2\theta)E_{x},\\
Q_{\perp}&=(\sigma_{zx}^{y}+\Delta_{zy}\cos^{2}\theta)E_{x}.
\end{aligned}
\ee
If $\Delta_{zy}\neq 0$, the spin polarization is  
perpendicular to $\hat{\vn{n}}$ only if $\hat{\vn{n}}$ is along
the $y$ or $z$-direction, otherwise spin polarization and direction of
spin current enclose the 
angle $\alpha=\arctan(Q_{\perp}/Q_{\Vert})\neq 90^{\circ}$, 
as shown in Fig.~1(b). 
It follows from Eq.~\eqref{eq_q_perp} that
$Q_{\perp}$ is zero at the angle
\bege\label{eq_theta_zero}
\theta_{0}=\arccos\sqrt{-\sigma_{zx}^{y}   /\Delta_{zy}  }
\ee
if $\sigma_{xy}^{z}$ and $\sigma_{zx}^{y}$ differ in sign.
At this angle $\theta_{0}$ the spin polarization and the spin current
are collinear. This is an interesting constellation, which cannot
occur in cubic systems.

The case of spin current in $x$-direction and
electric field $\vn{E}=(0,E\cos\theta,E\sin\theta)^{T}$ in the $yz$-plane is simply related 
to the previous one by a minus sign:
The components of the spin current with spin polarization parallel and
perpendicular to
the electric field $\vn{E}$ are given by
$Q_{\Vert}=\frac{1}{2}\Delta_{zy}\sin(2\theta)E$ and
$Q_{\perp}=-(\sigma_{zx}^{y}+\Delta_{zy}\cos^{2}\theta)E_{x}$,
respectively.
At the angle $\theta_{0}$, Eq.~\eqref{eq_theta_zero}, the spin polarization
and the electric field are collinear.
Thus, one can achieve collinearity of spin polarization and electric field,
or collinearity of spin polarization and direction of spin current if
$\sigma_{xy}^{z}$ and $\sigma_{zx}^{y}$ differ in sign.

If the electric field is applied along the $z$-axis, the same magnitude of the
spin current will be measured in all directions perpendicular to the $z$-axis,
since the $z$-axis exhibits 3-fold rotational symmetry.
The spin current in direction $\hat{\vn{n}}=(\cos\theta,\sin\theta,0)^{T}$
is in this case
\bege\label{eq_e_field_along_z_direction}
\hat{\vn{n}}\cdot\vn{Q}=
(
\sigma_{yz}^{x}\hat{\vn{s}}_{x}\sin\theta
-
\sigma_{zx}^{y}\hat{\vn{s}}_{y}\cos\theta
)E_{z}.
\ee
Symmetry requires $\sigma_{zx}^{y}=\sigma_{yz}^{x}$.
Consequently, the magnitude of the spin current is independent of $\theta$
and the spin polarization is perpendicular to both the electric field and
$\hat{\vn{n}}$. 

In the case of the hcp structure the conductivity vector 
and the spin current density, Eq.~\eqref{eq_define_spin_hall_conductivity_vector}, may be 
expressed in terms of the anisotropy as 
\bege
\label{eq_conductivity_vector_spin_polarization}
\begin{aligned}
\vht{\sigma}(\hat{\vn{S}})&=\sigma_{yz}^{x}\hat{\vn{S}}+(0,0,\Delta_{zy}S_{z})^{T},\\
\vn{Q}^{\hat{\vn{S}}}&=
\sigma_{yz}^{x}\vn{E}\times\hat{\vn{S}}+\Delta_{zy}S_{z}(E_{y},-E_{x},0)^{T}.
\end{aligned}
\ee
Hence, only two parameters, $\sigma_{yz}^{x}$ and $\Delta_{zy}$, suffice
to describe the SHE in hcp nonmagnetic metals.
This is a major difference to the anomalous Hall effect, where four
parameters are needed in the phenomenological 
expansion~\cite{roman_mokrousov_souza_2009} of 
the conductivity of hcp crystals
up to third order in the direction cosines, because the band energies
depend on the direction of magnetization.
Note that 
Eqns.~(\ref{eq_e_field_along_x}-\ref{eq_conductivity_vector_spin_polarization})
apply 
also to tetragonal metals.

For the sake of completeness we remark that
the analog 
of Eq.~\eqref{eq_define_anomalous_hall_conductivity_vector} 
and Eq.~\eqref{eq_define_spin_hall_conductivity_vector} runs 
$\vn{J}^{\text{OHE}}=\vn{E}\times\vht{\sigma}^{\text{OHE}}(\vn{B})$ in 
the case of the low field ordinary Hall effect (OHE) in paramagnets, where $\vn{B}$ is 
the magnetic induction and
$\vht{\sigma}^{\text{OHE}}(\vn{B})=\sigma_{yz}^{x}\vn{B}+(0,0,\Delta_{zy}B_{z})^{T}$
is the conductivity vector in hexagonal and tetragonal 
systems (cf Eq.~\eqref{eq_conductivity_vector_spin_polarization}).

In general there is an intrinsic (independent of impurities) and
an extrinsic (impurity-driven) contribution to the 
SHC (see Ref.~\cite{nagaosa_sinova_onoda_macdonald_ong_2010} 
and references therein for the origin of AHE, SHE is analogous).
In the first principles calculations of the SHC presented below 
we consider only the intrinsic~\cite{murakami_2003,sinova_2004,
Yao_Fang_2005,Guo_Murakami_Chen_Nagaosa} 
contribution to the SHC, which results from the virtual 
interband transitions in the
presence of an external electric field, and which  
may be 
written as a Kubo formula:
\bege\label{eq_kubo_for_she}
\begin{aligned}
\sigma_{ij}^{s}&=
e\hbar
\int \frac{d^{3}k}{(2\pi)^{3}}
\sum_{n=1}^{N_{\text{occ}}}
\Omega_{n}^{s}(\vn{k}),\\
\Omega_{n}^{s}(\vn{k})&=
2\Im\sum_{n\neq m}
\frac{
\langle \psi_{\vn{k}n}| Q_{i}^{s}|\psi_{\vn{k}m}\rangle
\langle \psi_{\vn{k}m} | v_{j} |\psi_{\vn{k}n}\rangle}{(\varepsilon_{n}-\varepsilon_{m})^{2}},
\end{aligned}
\ee
where $v_{j}$ is the $j$-component of the velocity operator,
$Q_{i}^{s}$ is the spatial $i$- and spin $s$-component of the
spin current density operator, $N_{\text{occ}}$ is the
number of occupied states, $|\psi_{\vn{k}n}\rangle$ is the Bloch function of
band $n$ at $k$-point $\vn{k}$ and $\varepsilon_{n}$ is its energy eigenvalue. 
If only the spin conserving part of SOI is taken into account the
spin current density operator may be written as 
$Q_{i}^{s}=\hbar/2 \,v_{i}\tau_{s}$.
Here, $\tau_{s}$ is a Pauli matrix used to express the $s$-component 
of the spin operator.
In order to treat the spin-nonconserving part of the SOI correctly 
we used the  
definition of the spin current density operator 
given in Ref.~\cite{Shi_Zhang_Xiao_Niu_2006}.

To disentangle the intrinsic and extrinsic contributions to the SHC
experimentally is still a challenge.
The anisotropy of the extrinsic AHE is expected to be much smaller 
than the one of the intrinsic AHE~\cite{roman_mokrousov_souza_2009}.
Since AHE and SHE are analogous concerning their intrinsic and extrinsic
mechanisms, also the anisotropy of the extrinsic SHE is expected to be 
small. 
Thus, the
direct comparison between the experimentally
measured anisotropy of the SHC and the one calculated theoretically
based on Eq.~\eqref{eq_kubo_for_she} allows to assess whether 
the first principles calculations predict the intrinsic contribution to
the SHC quantitatively correctly for a given system.
If quantitative agreement is found this
provides a strong justification
for the common procedure to attribute the difference between the experimentally
measured SHC and Eq.~\eqref{eq_kubo_for_she} to extrinsic effects, for 
the calculation of which
\textit{ab initio} methods have been 
developed recently~\cite{spin_projection_and_spin_current_density_ebert,
extrinsic_spin_hall_effect_from_first_principles,
spin_hall_angle_versus_spin_diffusion_length_tailored_by_impurities}. 
    
Our calculations of the intrinsic SHC, Eq.~\eqref{eq_kubo_for_she}, for the 
hcp metals
Sc, Ti, Zn, Y, Zr, Tc, Ru, Cd, La, Hf, Re and Os and for
antiferromagnetic
Cr
are based on the density functional
theory and were performed with the
full-potential linearized augmented-plane-wave (FLAPW) code {\tt FLEUR}~\cite{fleurcode}. 
The generalized gradient approximation of the exchange correlation potential,
a plane-wave 
cutoff of $K_{\text{max}}=3.7$~bohr$^{-1}$,
and the experimental lattice constants of the metals were chosen.  
In the case of Cr we neglected the spin-density wave and 
considered the antiferromagnetic structure with two atoms in the
unit cell and with the magnetic moments parallel and antiparallel
to the $z$-axis. 
A dense $k$-mesh is needed to
perform the Brillouin-zone integration in Eq.~\eqref{eq_kubo_for_she} accurately.
Consequently, we made use 
of Wannier 
interpolation~\cite{
calculation_anomalous_hall_conductivity_wannier_interpolation,
spectral_and_fermi_surface_properties_from_wannier_interpolation}
in order to reduce the computational cost. For this purpose we constructed
a set of 36 maximally localized FLAPW Wannier 
functions 
for each of the metals
using the Wannier90 code
(see Ref.~\cite{wannier90} and references therein)
and our 
interface~\cite{WannierPaper} between {\tt FLEUR} and
Wannier90. 

The resulting SHCs 
are shown in 
Fig.~2(a).
Except for Cd all metals studied in this work exhibit a large
anisotropy of SHE, which we expect to be clearly visible in
experiments.
Of particular interest are the hcp metals Sc, Ti and Ru, where the
sign of the conductivity changes as the spin polarization is rotated
from the $z$-axis into the $xy$-plane. As discussed before,
collinearity of the spin polarization and the electric field 
(of the spin polarization and the spin current) may be
achieved if the electric field (the spin current) lies in the
$yz$-plane
at the angle $\theta_{0}$, Eq.~\eqref{eq_theta_zero}, from the
$y$-axis.
To illustrate this we plot in
Fig.~2(b) the angle $\alpha$ enclosed by the
direction of the spin current and the direction of the spin polarization
as well as the SHCs associated with
$Q_{\Vert}$ and $Q_{\perp}$ (see Eq.~\eqref{eq_q_perp}) as a function of the
angle $\theta$ for Sc. 
The critical angles at which the perpendicular component of the spin polarization
vanishes are $\theta_{0}$=62.2$^{\circ}$, $\theta_{0}$=32.1$^{\circ}$, 
and $\theta_{0}$=19.1$^{\circ}$ for Sc, Ti, and Ru, respectively. 
In 
the case of Cr the SHE is anisotropic as the cubic
symmetry is broken by the staggered magnetization:
If the spin polarization of the spin current is
perpendicular to the staggered magnetization the SHC is larger by
a factor of 1.8 compared to the case of spin polarization parallel to
the staggered magnetization.

Generally, the integrand in Eq.~\eqref{eq_kubo_for_she} varies strongly as a
function of $\vn{k}$ and the entire Brillouin zone has to be
considered in the integration in order to reproduce the SHC
quantitatively correctly. This makes it hardly possible to interpret the
spin Hall conductivity in terms of a small number of virtual interband 
transitions. Even
the sign and order of magnitude of the SHC are difficult to predict based
on simple arguments. Recently, the variation of the sign of the 
Fermi-surface contribution to the SHC
along the 4$d$ and 5$d$ transition metal series has been attributed to
the variation of the sign of the spin orbit polarization on the Fermi
surface~\cite{giant_orbital_hall_kontani_2009}.
However, 
we find that the spin orbit polarization
does not change sign for Sc, Ti and Ru
while the SHC changes sign
as the spin polarization is rotated from the $z$-axis to the $y$-axis.

In conclusion, we have investigated the dependence of the SHE on the
directions of electric field and spin polarization. For the special
cases of hexagonal and tetragonal metals we derived the general
expression for the SHC vector. We predict that in hcp metals and 
antiferromagnetic Cr the SHE is strongly anisotropic. For Sr, Ti 
and Ru the anisotropy is particularly strong since the 
sign of the SHC depends on the orientation
of spin polarization. In this case
collinearity of spin polarization and electric field
(or spin polarization and spin current) can be achieved for special directions 
of the electric field (or of the spin current).

We gratefully acknowledge computing time on the supercomputers \mbox{JUGENE} 
and \mbox{JUROPA} 
at JSC and funding under the HGF-YIG programme VH-NG-513.

\end{document}